\documentclass{article}

\usepackage{microtype}
\usepackage{graphicx}
\usepackage{subfigure}
\usepackage{booktabs} 

\usepackage{hyperref}

\usepackage[accepted]{sysml2019}

\sysmltitlerunning{A Machine Learning Imaging Core using Separable FIR-IIR Filters}

\begin{document}

\twocolumn[
\sysmltitle{A Machine Learning Imaging Core using~Separable~FIR-IIR~Filters}



\sysmlsetsymbol{equal}{*}

\begin{sysmlauthorlist}
\sysmlauthor{Masayoshi Asama}{to}
\sysmlauthor{Leo F. Isikdogan}{to}
\sysmlauthor{Sushma Rao}{to}
\sysmlauthor{Bhavin V. Nayak}{to}
\sysmlauthor{Gilad Michael}{to}
\end{sysmlauthorlist}

\sysmlaffiliation{to}{Intel Corporation, Santa Clara, CA}

\sysmlcorrespondingauthor{Leo F. Isikdogan}{leo.f.isikdogan@intel.com}
\sysmlcorrespondingauthor{Masayoshi Asama}{masayoshi.asama@intel.com}

\sysmlkeywords{Machine Learning, SysML}

\vskip 0.3in

\begin{abstract}
We propose fixed-function neural network hardware that is designed to perform pixel-to-pixel image transformations in a highly efficient way. We use a fully trainable, fixed-topology neural network to build a model that can perform a wide variety of image processing tasks. Our model uses compressed skip lines and hybrid FIR-IIR blocks to reduce the latency and hardware footprint. Our proposed Machine Learning Imaging Core, dubbed MagIC, uses a silicon area of $\sim$3mm$^2$ (in TSMC 16nm), which is orders of magnitude smaller than a comparable pixel-wise dense prediction model. MagIC requires no DDR bandwidth, no SRAM, and practically no external memory. Each MagIC core consumes 56mW (215 mW max power) at 500MHz and achieves an energy-efficient throughput of 23TOPS/W/mm$^2$. MagIC can be used as a multi-purpose image processing block in an imaging pipeline, approximating compute-heavy image processing applications, such as image deblurring, denoising, and colorization, within the power and silicon area limits of mobile devices.

\end{abstract}
]



\printAffiliationsAndNotice{}  

\section{Introduction}

Many convolutional neural network (CNN) architectures that make dense, pixel-wise predictions, such as FCN~\cite{long2015fully}, U-Net~\cite{ronneberger2015u}, and their variants, use very long skip lines. Those skip lines are crucial for recovering the details lost during downsampling. However, hardware implementations of those networks require a large memory to hold those skip lines. The skip lines are often stored in external memory such as SRAM or DDR, which dramatically increases the cost in terms of silicon area footprint. Furthermore, the skip connections cause additional latency, which would hinder real-time applications of the system.

In real-time imaging systems, images are acquired line-by-line by the raster scan order. Therefore, an efficient hardware implementation of a CNN that runs in such a system needs to be fully-pipelined. However, it is a challenging task to implement a CNN topology that has long skip connections in fully-pipelined hardware in a cost-effective way. The main reason is that skip lines need to compensate for all vertical delays of the entire network.

\begin{figure}[t]
\centering
\vspace{5pt}
\includegraphics[width=1.0\linewidth]{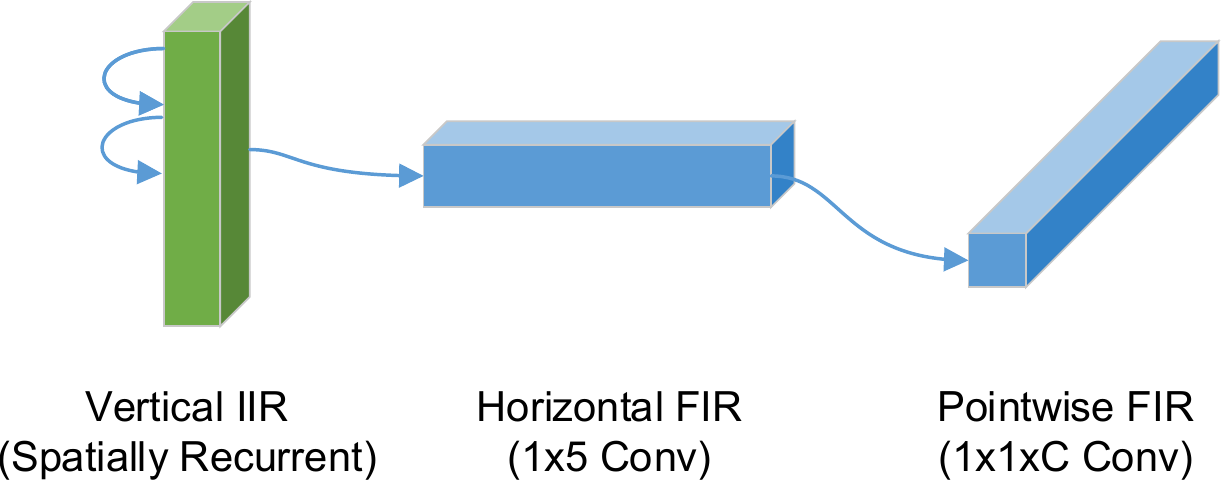}
\vspace{5pt}
\caption{We break down a convolution layer into separable convolutions in all three dimensions and replace the vertical component with an IIR filter. The vertical IIR reduces the vertical delay in a line-based hardware implementation, leading to significant savings in silicon area.}
\label{fig:hybrid_iir_fir}
\end{figure}

In a line based system, the vertical line delay is accumulated in every convolution layer. For example, a 3x3 spatial window would cause a 1-line delay, whereas two consecutive 3x3 convolutions would result in a 2-line delay. The problem with the long skip lines is that once the data on one end of the skip line is generated, it needs to be held in memory until the data in the receiving end of the skip connection is ready. The more layers a connection skips over, the more lines need to be kept in memory. Therefore, the size of the total memory required increases with the length of the skip line. The memory requirements for the skip lines can aggregate quickly and become a significant contributor to the total silicon area needed to implement the network. Moreover, the latency caused by the accumulated line delay can also be problematic in latency-sensitive applications such as autonomous driving systems.

A naive way to reduce the vertical delay in such a model would be to reduce the number of pooling and convolution layers. However, this would reduce the receptive field, which dictates the size of the area in the input that can affect a single pixel at the output. The more layers and scales a neural network has and the bigger kernels are, the larger the receptive field becomes.

For image processing tasks, a large receptive field is needed to be able to process medium and low spatial frequencies. Furthermore, many tasks that lie at the intersection of image processing and computer vision need a sizeable receptive field to be able to make context-aware decisions. For example, an image colorization model would need semantic cues that span a large portion of its input to detect the sky and process the pixels accordingly.

We propose a hardware-friendly neural network topology that maintains a large receptive field without producing large vertical delay lines. Our method significantly reduces the memory requirements while reducing end-to-end latency by replacing some of the finite impulse response (FIR) filters with infinite impulse response (IIR) filters and compressing the skip lines. We use this model to implement a Machine Learning Imaging Core (MagIC) as fixed-function hardware having configurable parameters.

MagIC can potentially work in multiple locations in an imaging pipeline to implement or complement certain features in pre-processing, post-processing, or anywhere in between in the pipe. For example, MagIC can 
\begin{itemize}
\itemsep0em
  \item improve the image quality by learning a mapping between the outputs of low-cost and high-end image signal processors (ISPs);
  \item approximate compute-heavy image processing operators, such as denoising and deblurring algorithms;
  \item recover missing color information from the context, such as converting RCCC (Red/Clear) images used in advanced driver assistance systems to full color RGB images;
  \item process single or stereo camera input to create depth maps;
  \item demosaic non-traditional color filter array images, such as hybrid RGB-IR and spatially varying exposures.
\end{itemize}

The flexibility of MagIC would help address customer-specific requests without changing the underlying ISP hardware. This would help create low-cost, yet powerful imaging systems.

\section{Related Work}
Our machine learning imaging core can be considered a type of multi-purpose image signal processor. Image signal processors (ISPs) typically implement a fully pipelined image processing architecture that makes use of line buffers to store all intermediate data between different stages of processing. This architectural pattern provides highly efficient image processing pipelines that achieve high throughput. Prior work on building optimized imaging pipelines, such as Darkroom~\cite{hegarty2014darkroom} and FlexISP~\cite{heide2014flexisp}, mainly focused on implementing particular image processing algorithms efficiently on hardware. Although defining image processing operators as fixed-function ASIC blocks generally improves the performance of a system, many algorithms can be still too complex to run on low power environments.

Recent work~\cite{chen2017fast,gharbi2017deep} showed that many sophisticated image processing operators could be efficiently approximated using convolutional neural networks. Practically, a U-Net-like~\cite{ronneberger2015u} neural network topology that outputs pixel-wise labels can approximate virtually any image processing operator, although implementing the U-Net as-is in hardware would be very costly.

Our work implements a convolutional neural network as a fully-pipelined, multi-purpose imaging block. Prior work on fully-pipelined hardware implementations of neural networks~\cite{visionisp,whatmough2019fixynn} focused on computer vision tasks. For example, the VisionISP pipeline~\cite{wu2019trainable} used a trainable vision scaler (TVS), which implemented a shallow convolutional neural network as in ISP block to perform vision-aware image downscaling. It is indeed technically possible to train TVS to approximate some image processing operators. However, TVS was designed to be a pre-processor for computer vision tasks and would not have sufficient model capacity as a standalone module to perform advanced imaging tasks, such as image colorization~\cite{zhang2016colorful} and advanced image deblurring~\cite{nah2017deep}.

Another fixed-topology neural network hardware, called FixyNN~\cite{whatmough2019fixynn}, froze the first layers of a MobileNet~\cite{howard2017mobilenets} model and implemented it as a fixed feature extractor. Although this approach worked well for image classification, using such feature extractor as an image processing block would be challenging. Images in an ISP pipeline can have very different characteristics, such as having different lens shading and noise profiles. The location of the block in a pipe can also change the properties of the input. For example, images would look different before and after denoising and sharpening. A model would need to have configurable parameters to have the flexibility to adapt to different types of inputs. Furthermore, the amount of downscaling done in image classification feature extractors makes them unfeasible for learning pixel-to-pixel transformations.

\begin{figure*}[t]
\centering
\vspace{5pt}
\includegraphics[width=1.0\linewidth]{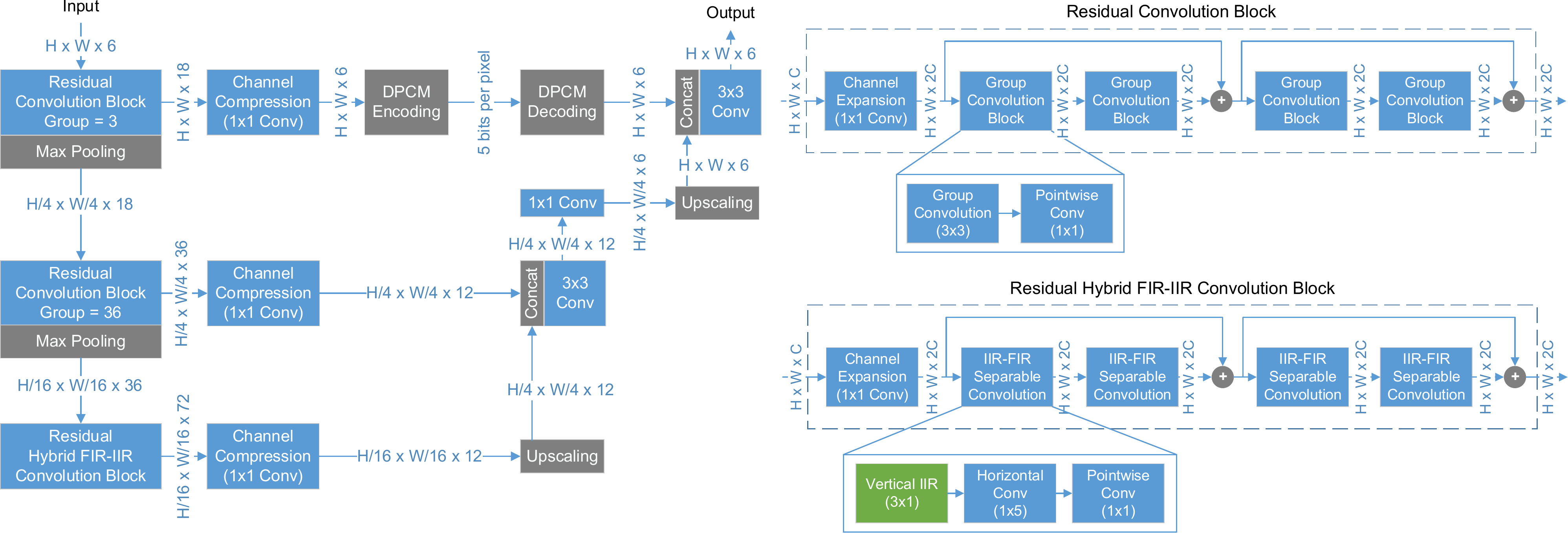}
\vspace{5pt}
\caption{Architecture of the network that we use to implement our machine learning imaging core. Our model uses compressed skip lines and hybrid FIR-IIR blocks to reduce the amount of data that needs to be line-buffered in fully-pipelined hardware.}
\label{fig:model_architecture}
\end{figure*}

\section{Model Architecture}
In this section, we describe the topology of the network that we use to implement our machine learning imaging core. We explain the design choices we made to build a hardware-efficient model. 

\subsection{Skip-connection Compression}
The macro architecture of our model is a typical, U-Net style~\cite{ronneberger2015u}, encoder-decoder network that has skip connections between layers at the same spatial resolution (Figure \ref{fig:model_architecture}). One challenge about implementing a U-Net-like model is the memory cost associated with the long skip connections. Indeed, it is possible to remove the skip lines altogether to save hardware area. However, removing even only the longest skip connection results in a drastic drop in output image quality as those skip lines help recover the spatial granularity that is lost after downscaling layers.

Instead of removing the skip connections, we compress the data carried over them to reduce the memory requirement of the model. First, we reduce the number of channels on a skip line buffer by using point-wise convolutions, acting as trainable linear projection layers. After this channel-wise compression, we also use differential pulse code modulation (DPCM)~\cite{cutler1952differential} to reduce the number of bits needed to store each pixel on the buffer. We use DPCM compression only on the longest skip line, where the silicon area cost of the DPCM encoder and decoder are negligible as compared to the cost of the skip line buffers.

Overall, compressing the skip lines allows us to use only internal memories and no external memory for the entire inference operation.

\subsection{Separable FIR-IIR Filters}
The concept of separable convolutions is commonly used to design efficient neural network architectures, particularly in the form of depthwise-separable convolutions~\cite{howard2017mobilenets,chollet2017xception,sandler2018mobilenetv2}. Depthwise-separable convolutions replace a $K \times K \times C_{in} \times C_{out}$ convolution with $K \times K$ convolutions for each input channel $C_{in}$, followed by a point-wise $1 \times 1 \times C_{in} \times C_{out}$ convolution. Depthwise separation usually leads to significant savings in the number parameters since $K \times K + C_{out}$ is typically much smaller than $K \times K \times C_{out}$. It is possible to take this kernel separability one step further and separate a $K \times K$ filter spatially as $K \times 1$ and $1 \times K$ filters. This type of spatial separation is not commonly used in modern convolutional neural network architectures, since spatial separation does not reduce the number of parameters significantly enough for small kernel sizes. However, spatial separability would still provide benefits when the kernels are large, and the cost of horizontal and vertical convolutions are not the same.

In a line-based system, the cost of vertical convolutions can be disproportionally high due to the number of lines that need to be buffered before the convolution for a given window can be computed. For example, a $1 \times 5$ convolution would need only 4 elements to be buffered, whereas a $5 \times 1$ convolution would need 4 lines of data to be buffered. We address this problem by replacing the vertical convolutions with IIR filters (Figure \ref{fig:hybrid_iir_fir}).

Using an IIR filter in the vertical direction can approximate a convolution without producing vertical delay lines. We use a first-order IIR to approximate a vertical (Nx1) convolution. We implement this operator as a spatially-recurrent neural network cell as:
\begin{equation}
    h[t] = h[t-1]\cdot w_1 + x[t-1]\cdot w_2 + x[t]\cdot w_3
\end{equation}
where $x$ is the input, $h$ is the output, $w$ stands for the trainable weights, and $t$ indicates the spatial position in the vertical axis rather than time.

Recurrent modules are typically used to train machine learning models on time series data. In our case, they are used to summarize pixels in the vertical direction. Unlike fixed-window convolutions, a recurrent module can start processing its input as the pixels arrive line by line without having to buffer the lines that are spanned by the fixed-sized window. Therefore, using a recurrent module in the vertical direction reduces the time distance between the input and the output of the model.

The recurrent module we use approximates a simple column-convolution and is not expected to remember long term dependencies. Therefore, it does not use sophisticated gating mechanisms as long short-term memory (LSTM) or gated recurrent unit (GRU) modules do.

We use the spatially recurrent modules to define hybrid FIR-IIR blocks that replace more expensive convolution blocks. The hybrid FIR-IIR blocks use 3-way (horizontal, vertical, and depthwise) separable convolutions, where IIR filters approximate the vertical components.

We use the hybrid FIR-IIR blocks only in the coarsest-scale (bottleneck) layers, where the impact of convolution on the overall vertical delay in the system is the largest. For example, a 3x3 filter in the bottleneck would cause a 16-line delay as compared to a 1-line delay in the first layer. Furthermore, IIR filters are known to handle low frequencies well. This characteristic makes hybrid IIR-FIR filters well suited for the bottleneck layer, which processes low-frequency features. Since the height of the feature maps at the bottleneck layer is reasonably small, our model does not suffer from exploding or vanishing gradient problems.

In the other scales, we use convolution blocks that consist of group convolutions followed by pointwise convolutions, where the number of groups is tuned to find a balance between hardware cost and the quality of the produced images. The convolution blocks in the first scale have 3 groups of convolutions. The second scale convolutions have their number of groups equal to the number of their input channels, which makes them depthwise separable convolutions.

\subsection{Other Design Choices}
Our model inputs and outputs up to 6-channel images. Those six channels can be used to process two RGB images captured by a pair of cameras, two consecutive frames captured by a single camera, or a 6-band multispectral image acquired by single or multiple sensors. This design choice opens up possibilities for a variety of applications that involve stereo depth, sensor fusion, multispectral imagery, and temporal processing.

The encoder part of our model uses ResNet-like~\cite{resnet} residual connections. Unlike the long skip lines, those residual connections have a minimal cost in hardware. Using residual connections results in easier-to-train models, stabilized the training, and improved the consistency in results.

The max-pooling layers in our model use a stride of 4 to be able to cover a broader range of scales using fewer layers. Our empirical results showed that given the same number of max-pooling layers, using a stride of 4 produces higher-quality outputs than using a stride of 2. Using twice as many encoder blocks followed by max-pooling layers having a stride of two does further improve the results but at the cost of increased hardware area. Reducing the depth of the model using 4x4 max-pooling layers helped us reduce the hardware footprint and design a neural network topology for very small silicon area budgets ($\sim$3mm$^2$). For smaller area budgets, we also provide the option of approximating the multiplications in the model with low-cost shift-sum multipliers~\cite{oron2017instruction}, which further reduces the hardware cost.

\begin{table}[t]
\centering
\vspace{5pt}
\includegraphics[width=0.98\linewidth]{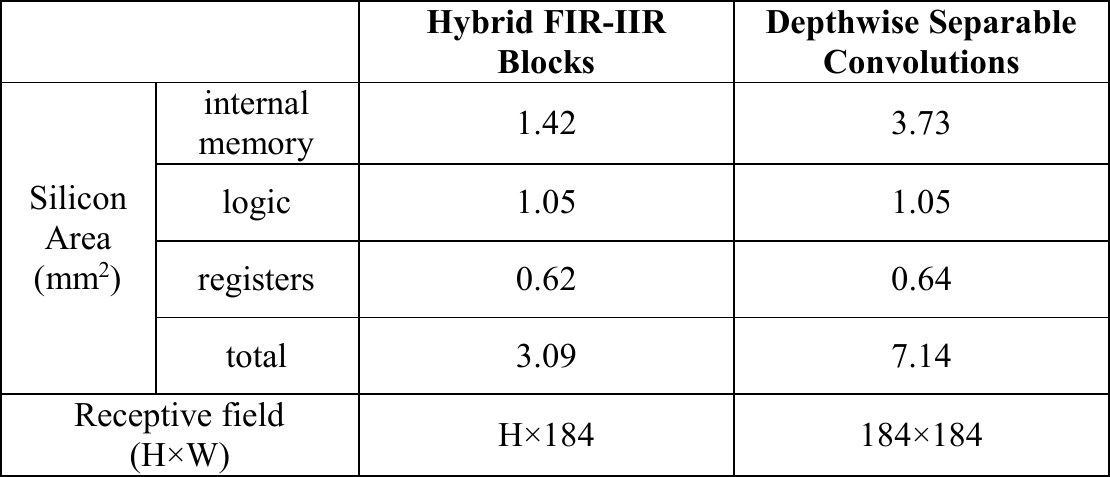}
\vspace{3pt}
\caption{Silicon area cost and receptive field of using Hybrid FIR-IIR blocks at the bottleneck layer as compared to using depthwise separable convolutions. Hybrid FIR-IIR blocks lead to significant savings in the hardware footprint associated with internal memory.}
\label{tab:hardware_area}
\end{table}

\begin{figure*}[t]
\centering
\vspace{5pt}
\includegraphics[width=1.0\linewidth]{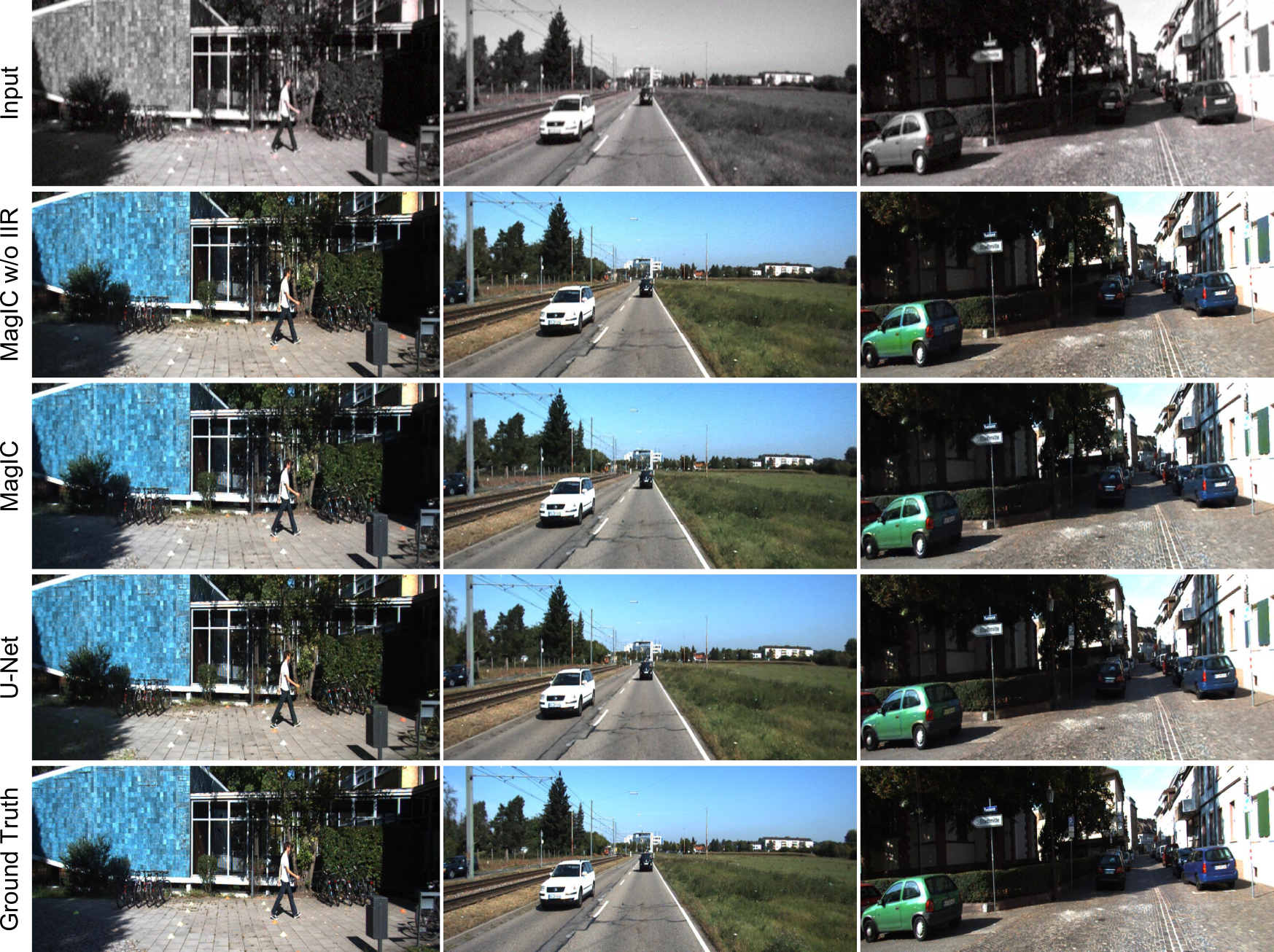}
\caption{Qualitative comparison of MagIC and U-Net. Given distorted Red/Clear images, the models denoise and deblur the images while restoring the missing color information to produce clean, sharp, full-color RGB images. From top to down: input images, MagIC using traditional depthwise separable convolutions, MagIC using our proposed hybrid FIR-IIR blocks, U-Net, and ground truth. Outputs of the two variants of MagIC are virtually the same although the one using the hybrid FIR-IIR blocks has a much smaller hardware footprint.}
\label{fig:results}
\end{figure*}

\begin{table}[t]
\centering
\vspace{5pt}
\includegraphics[width=0.98\linewidth]{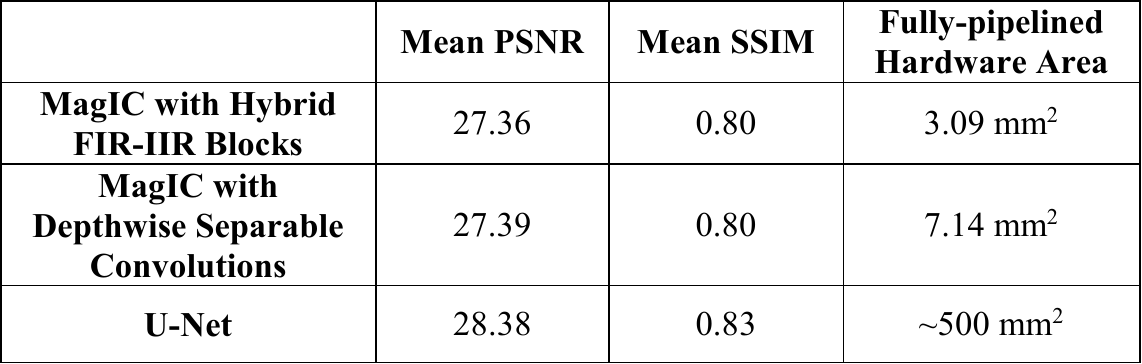}
\vspace{3pt}
\caption{Comparison of variants of MagIC and a fully-blown U-Net model in terms of PSNR and SSIM on the test set as well as silicon area when the network topologies are modeled as line-buffered hardware using TSMC 16nm technology.}
\label{tab:results}
\end{table}

\begin{figure}[t]
\centering
\vspace{5pt}
\includegraphics[width=0.95\linewidth]{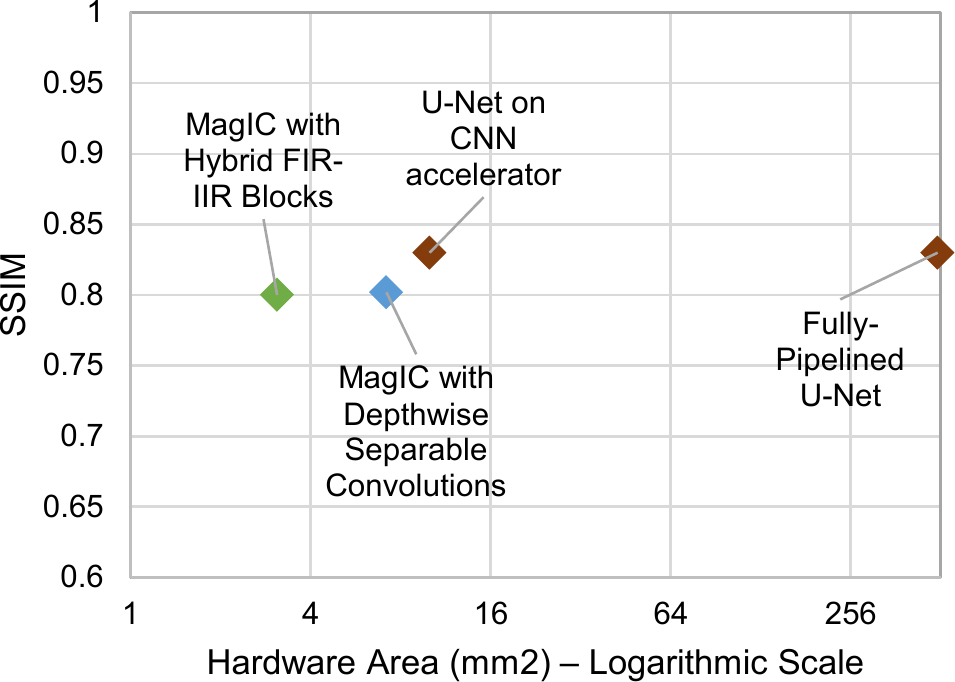}
\vspace{5pt}
\caption{Comparison of MagIC with and without the hybrid FIR-IIR blocks, U-Net implemented as fully-pipelined hardware, and U-Net running on a CNN accelerator.}
\label{fig:area_vs_ssim}
\end{figure}

In our area calculations, we assumed the input to be $1920\times 1080$ pixels at 30 frames per second. Both the input resolution and the frame rate impacts the total silicon area needed to implement our proposed hardware. Specifically, the input image size affects the internal memory area, whereas the frame rate impacts the logic area (Table~\ref{tab:hardware_area}). The area for the memory scales linearly with the image width since the images are kept on line buffers. Wider inputs require larger line buffers to accommodate the input, output, and intermediate feature maps throughout the pipeline. The logic area also scales approximately linearly with the frame rate except for frame rates lower than 30 fps. For example, doubling the frame rate would require twice as many pixels to be processed in one cycle, resulting in approximately twice as large logic area. However, reducing the frame rate by half would not lead to proportional savings in the logic area due to overheads and hardware inefficiencies.

We calculated the area by modeling our proposed hardware using the TSMC 16nm process technology. However, the actual silicon area of the hardware can be made smaller. The area would be further reduced using more contemporary manufacturing process technologies, such as the Intel 10nm technology.

\section{Results}
We evaluated our proposed model on a combined denoising, deblurring, and coloring task on the KITTI dataset~\cite{kitti}, emulating an imperfect front view camera on a vehicle. This exemplary task encompassed both low-level image processing operations such as denoising and sharpening as well as higher-level tasks such as inferring the color of an object given texture and context.

We used images in the KITTI dataset as reference images and generated distorted versions of those images. First, we applied an approximate color space conversion to the input images to emulate RCCC (Red/Clear) sensors. RCCC sensors use clear filters instead of the blue and green filters and are typically designed for automotive use. We converted RGB images to RCC images by replacing the green and blue channels by grayscale image intensity (Y) channels. The Y-channel approximated the clear (C) channels in an RCCC image. Then, we blurred the images using a random Gaussian blur operator to simulate a point spread function. Finally, we added Gaussian noise to each channel independently, in the linear domain, at two different scales. We used a noise variance that correlated with signal intensity, simulating the noise profile of a real sensor. We separated a portion of the resulting image pairs for test and used the rest for training.

We trained MagIC to reconstruct the reference images given the distorted images as input, learning to denoise and deblur the distorted images while restoring the missing color information in the RCCC input. To measure the impact of the hybrid FIR-IIR blocks, we also trained a variant of MagIC that used depthwise separable convolutions instead of separable FIR-IIR filters. Using the hybrid FIR-IIR blocks significantly reduced the hardware footprint (Table~\ref{tab:hardware_area}) without having a negative impact on qualitative (Figure~\ref{fig:results}) and quantitative results (Table~\ref{tab:results}).

We also compared both variants to a fully-blown U-Net~\cite{ronneberger2015u} model trained using the same setup. As expected, the U-Net model had higher PSNR and SSIM scores than MagIC, given its orders of magnitude larger model capacity. However, MagIC was able to achieve an image quality close to the reference U-Net model within an area budget of only $\sim$3mm$^2$, modeled with TSMC 16nm technology. If we were to implement U-Net as-is using the same fully-pipelined hardware architecture, the overall footprint of the resulting hardware would be over 500mm$^2$. It would indeed be more feasible to implement U-Net using a generic CNN accelerator within an area budget of $\sim$10mm$^2$ rather than a fully-pipelined hardware block (Figure~\ref{fig:area_vs_ssim}). However, using a generic accelerator would result in lower utilization rate, lower frame rate, and over $8\times$ power consumption, while still being over 3$\times$ larger than our solution.

\section{Conclusion}
We described a low-cost machine learning imaging core that used a fixed-topology neural network to process images in a multitude of ways. We used a bag of tricks to minimize the silicon area needed to implement our model. We used 3-way separable convolutions and approximated the convolutions in the vertical direction using infinite impulse response (IIR) filters. This approximation significantly reduced the silicon area needed to implement the underlying model in hardware. Our proposed hybrid FIR-IIR blocks not only reduced the latency but also increased the receptive field of the model, improving the contextual coherence of the results. We further reduced the cost of our proposed system by compressing skip lines and carefully designing the topology of our model. Finally, we showed that our proposed hardware was able to perform both low-level and high-level imaging tasks concurrently, within a silicon area budget of only $\sim$3mm$^2$.

In this paper, we focused on optimizing a U-Net-like neural network architecture to perform pixel-to-pixel image processing. We believe that our proposed methods have the potential to be useful in numerous applications beyond image processing. For example, the separable FIR-IIR filters can help shallow image classification models capture the context information better. As future work, it would be interesting to study how the concepts presented in this paper would generalize to a broader range of applications.

\bibliographystyle{sysml2019}

\vfill
\section*{Disclaimer}
No license (express or implied, by estoppel or otherwise) to any intellectual property rights is granted by this document. This document contains information on products, services and/or processes in development. All information provided here is
subject to change without notice. Intel and the Intel logo are trademarks of Intel Corporation in the U.S. and/or other
countries. \\ {\small\textcopyright}  Intel Corporation.

\end{document}